\documentclass[aps,floats,showpacs,superscriptaddress,twocolumn]{revtex4-1}
\usepackage{graphicx}
\usepackage{epstopdf}
\usepackage{amssymb}
\usepackage{amsmath}
\usepackage{mathtools}

\begin{document}
\title{Polarized neutron scattering on HYSPEC: the HYbrid SPECtrometer at SNS}

\author{Igor A Zaliznyak}
\email{zaliznyak@bnl.gov}
\address{CMPMSD, %Condensed Matter Physics and Materials Science Division,
 Brookhaven National Laboratory, Upton, NY 11973 USA}
\author{Andrei T. Savici}
\author{V. Ovidiu Garlea}
\author{Barry Winn}
\address{NScD, %Neutron Sciences Division,
 Oak Ridge National Laboratory, Oak Ridge, TN 37831 USA}
\author{Uwe Filges}
\address{Paul Scherrer Institut, Laboratory for Scientific Developments and novel Materials, Villigen PSI, Switzerland}
\author{John Schneeloch}
\author{John M. Tranquada}
\author{Genda Gu}
\author{Aifeng Wang}
\author{Cedomir Petrovic}
\address{CMPMSD, %Condensed Matter Physics and Materials Science Division,
 Brookhaven National Laboratory, Upton, NY 11973 USA}

%\ead{zaliznyak@bnl.gov}

\begin{abstract}
We describe some of the first polarized neutron scattering measurements performed at HYSPEC \cite{Zaliznyak_PhysicaB2005,ShapiroZaliznyak_PhysicaB2005,Winn2015,HYSPEC_website} spectrometer at the Spallation Neutron Source, Oak Ridge National Laboratory. We discuss details of the instrument setup and the experimental procedures in the mode with full polarization analysis. Examples of polarized neutron diffraction and polarized inelastic neutron data obtained on single crystal samples are presented.
\end{abstract}

\maketitle

\section{Introduction}

Neutron polarization analysis (NPA) \cite{Moon1969,Maleev_JETP1961,IzyumovMaleev_JETP1962,Blume_PhysRev1963} provides an unprecedented range of capabilities, both for developing new neutron scattering techniques \cite{PappasEhlersMezei2006,Regnault2006,Maleev_Uspekhi2002}, and for elastic and inelastic neutron scattering studies of novel magnetic states and excitations. Polarized neutron measurements allow one to identify the directions of magnetic moments and their fluctuations, distinguish between the structural and magnetic scattering features, and separate the magnetic and the non-magnetic background. There is a number of recent examples, which show how NPA enables disentangling the nature of complex magnetic states \cite{Maleev_Uspekhi2002,Schweizer2006,Brown_PhysicaB2001}, understanding the nature of magnetic excitations in quantum magnets \cite{Raymond_PRL1999,Lake_PRL2000,Regnault_PhysicaB2004,Lake_PRB2005,Lake_NatMat_2005}, cuprate \cite{Fong_etal_Keimer_PRB2000,Headings_etal_PRB2011} and heavy fermion \cite{RaymondLapertot_PRL2015} superconductors, and probing novel types of magnetic dynamics, such as chirality fluctuations \cite{Maleev_Uspekhi2002,Plakhty_EPL1999,Plakhty_PRB2001,Lorenzo_etal_PRB2007}. Recent discoveries of unusual temperature-induced dynamical magnetism \cite{Zaliznyak_PRL2011,Soh_etal_McQueeney_PRL2013} and unexpected, ``forbidden'' structural modes in the iron pnictides and chalcogenides \cite{FobesZaliznyak_PRL2014,FobesZaliznyak_PRB2016} emphasize the special significance of polarization analysis for strongly-correlated electron systems and unconventional superconductors. Only via NPA can these features be uniquely identified; an example of such a polarized neutron measurement performed on HYSPEC will be presented below.

Traditionally, polarized inelastic neutron experiments were carried out using triple-axis spectrometers (TAS) \cite{Moon1969,Regnault2006,TASBook,ZaliznyakLee_MNSChapter}, with narrow and collimated beams. The corresponding instrument geometries are well suited for employing diverse neutron beam polarization methods, such as Bragg reflection from a Heusler crystal, or total reflection from a supermirror, or transmission through a reasonably small vessel filled with polarized $^3$He. However, much of the recent progress in neutron spectroscopic techniques has been associated with the development of time-of-flight (TOF) direct geometry chopper spectrometers (DGCS) at spallation sources, which feature large detector arrays covering up to $\approx 2-3$~sr solid angle \cite{Stone_RSI2014,Ewings2016}. Such wide-open geometry is in conflict with the requirements of most traditional beam polarization and NPA methods, which is further exacerbated by the usage of broad neutron energy bands, including thermal and high-energy epi-thermal neutrons.

\begin{figure}[b]
%\begin{center}
\includegraphics[width=0.49\textwidth]{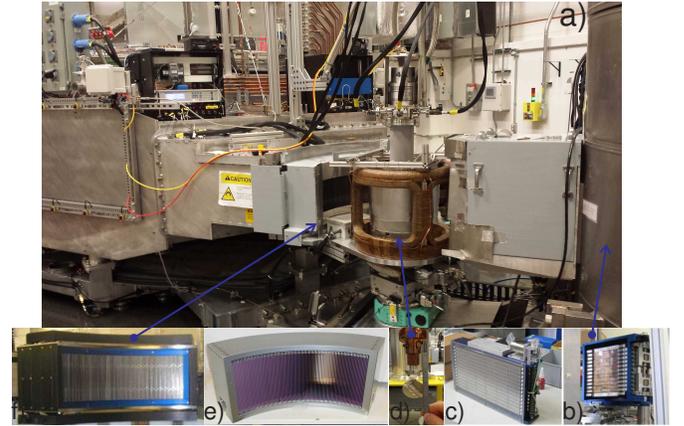}
%\end{center}
\caption{HYSPEC instrument configured for polarized beam measurements (a) and the beamline components involved in the setup, (b)-(f). The incident beam is polarized by the vertically focussing magnetized Heusler crystal array, (b), which by the means of an automated vertical translation stage is interchangeable with the co-axially mounted pyrolytic graphite (PG) monochromator, (c), used for the un-polarized measurements. The coils mounted at the sample table provide adiabatic guide field of up to $\sim 50$~Gauss at the sample, (d), whose orientation can be freely tuned. The radial collimator, (e), used for the un-polarized measurements is replaced by the identically sized supermirror analyzer, (f).}
\label{fig:InstrumentView}
\end{figure}

One approach to meeting the requirements of wide-angle, broad-band scattered beam polarization analysis relies on developing wide-angle $^3$He cells, which can be placed between the sample and the detectors. Such a method can provide high polarization efficiency and good transmission in a broad energy range, but at the same time has several significant deficiencies. Firstly, the beam polarization depends on the $^3$He polarization, which typically is time-dependent and has to be periodically regenerated. For data acquired in the event mode, which is now standard on modern instruments at spallation sources, this can be adequately treated in the data analysis software, although a disadvantage of the labor-intensive experimental procedure associated with maintaining the $^3$He polarization remains. A more significant problem stems from the extreme sensitivity of the $^3$He system to magnetic fields, which restricts the usage of magnetic fields in the sample environment and typically limits the NPA to only a single (vertical) guide field direction, thus markedly reducing its potential. An alternative approach, which is largely free of these limitations, relies on developing a wide-angle supermirror polarizer (SP) system \cite{Zaliznyak_PhysicaB2005,Stewart2009,PINS2006,Filges2006,Filges2012}. Such systems are low-maintenance, easy to operate, and are weakly sensitive to magnetic fields, thus allowing the XYZ-polarization analysis \cite{Scharpf1993,Ehlers2015}. However, they have a strongly energy-dependent transmission, which is very low in the thermal to epithermal neutron energy range, thus limiting significantly the neutron energy bandwidth amenable to NPA.

The HYSPEC instrument \cite{Zaliznyak_PhysicaB2005,ShapiroZaliznyak_PhysicaB2005,Winn2015} at the Spallation Neutron Source (SNS), Oak Ridge National Laboratory, was designed with the goal of implementing wide-angle polarization analysis. HYSPEC's design accommodates both the $^3$He polarization analyzer (PA) option, which is still at the development stage, and the supermirror PA, which was delivered to SNS in July of 2015 and is now in service. The supermirror polarization analyzer has been designed and built by the Paul-Scherrer Institute (PSI), where the associated unique supermirror manufacturing and assemblage technology was also developed. HYSPEC operation in the fully polarized mode became available since the end of 2015, and here we report on some of the first polarized neutron experiments on HYSPEC with full NPA.

\section{General overview of HYSPEC design and specifications}

HYSPEC is a high-intensity, medium-resolution thermal neutron TOF spectrometer located at SNS beam line 14B viewing a cryogenic coupled-hydrogen moderator \cite{Winn2015,HYSPEC_website}. The neutron beam is delivered by a $\approx 38.7$~m long, 4~cm wide by 15~cm tall neutron guide (initial $\approx 4$~m section expands the beam height from 13.2 cm to 15 cm) with $m = 3(2)$ outer(inner) supermirror coating. The $\approx 24$~m long central section of the guide is curved ($R = 2558.7$~m), and offsets the guide exit by $\approx 16$~cm, well beyond the line of sight of the source. The resulting guide cutoff is at $E_i \approx 120$~meV, with a factor $\approx 2$ beam attenuation at $E_i = 90$~meV.

The thermal neutron beam delivered by the guide is Bragg-reflected by a vertically focussing crystal array located $\approx 39$~m from the source and $1.8$~m from the sample. Focussing the 15 cm tall beam increases neutron flux on a typical 2~cm tall sample by a factor 3 to 6, at the expense of the vertical wave vector resolution. The latter is determined by the angular view of the 15~cm tall beam incident on the crystal array, from the sample position, which is $\approx 4.8^\circ$, or about 1/3 of the detector height. The beam is cleaned of contamination by a set of disk choppers and a T$_0$ chopper similar to those on ARCS and SEQUOIA \cite{Stone_RSI2014}, and is monochromatized by a short-blade (``rotating collimator'') Fermi chopper. The latter is located $1.81$~m upstream of the vertically focussing crystal array, which is placed in a rotating drum shield and is commonly called the ``monochromator'' because its design is similar to that of a focussing crystal monochromator of a triple axis spectrometer \cite{TASBook}. The Fermi chopper rotation frequency can be set between 30~Hz and 420~Hz, in 30~Hz increments, which yields the elastic energy resolution full width at half maximum (FWHM) from $\approx 10\%$ down to $\approx 2.3\%$ of E$_i$. The latter is roughly equivalent to halving the resolution FWHM of the triple axis spectrometer at the same E$_i$.

The sample stage and the detector vessel on HYSPEC move in registry with the rotating drum shield when the incident neutron energy is changed. The sample-monochromator optics is similar in design to a that of a triple-axis spectrometer. The open air sample stage, which provides sample rotation, translation, and tilts is attached with a rigid arm to the monochromator drum shield and moves on air pads when the drum rotates. The collimators, slits, flippers, and other auxiliary primary beam line components are mounted on the optic stage attached to the drum. The entrance to the scattered neutron flight path in the Ar-filled detector vessel is defined by a radial collimator (in an unpolarized mode), at about 50~cm from the sample axis. This permits the use of a variety of sample environments.

The scattered neutron energy is determined by the time-of-flight between the sample and the detectors. The detector bank is a cylindrical array of 20 8-packs of 1.2~m long, $1''$ diameter $^3$He linear position sensitive (128-pixel) detector tubes vertically mounted on the back wall of the detector vessel. Detectors are positioned coaxially with the sample rotation axis, at a nominal 4.5~m distance from it. The HYSPEC detector bank covers $60^\circ$ horizontal by $15^\circ$ vertical, for a nominal 0.25~sr solid angle (the actual physical coverage by detector pixels is 0.226 sr \cite{Stone_RSI2014}). The detector vessel is attached with a rigid arm to the sample stage and can independently rotate around the sample axis to cover scattering angles from 0$^\circ$ up to 135$^\circ$ \cite{Winn2015,HYSPEC_website}.

\section{HYSPEC setup and the measurement procedure in the polarized beam mode}

\begin{figure}
%\begin{center}
\includegraphics[width=0.49\textwidth]{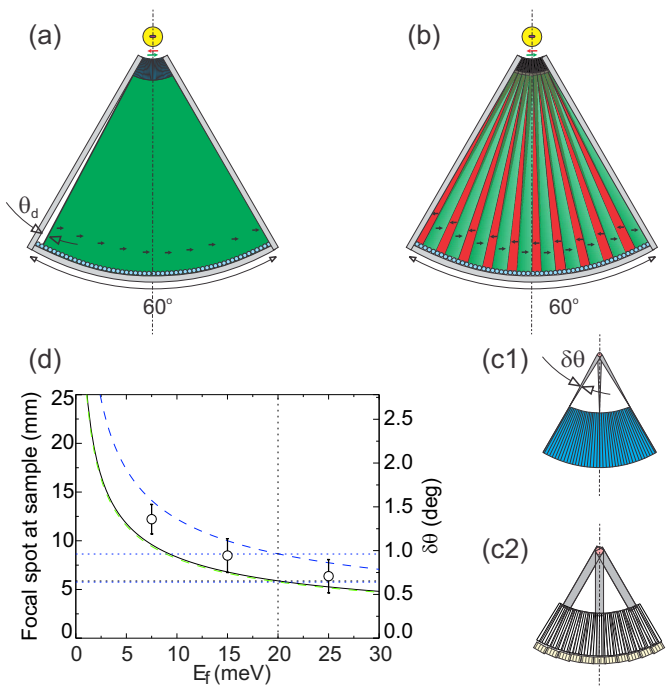}
%\end{center}
\caption{Schematics of the scattered neutron polarization analysis on HYSPEC: (a) as currently operational, with a radial supermirror polarizer that transmits the deflected neutrons with one polarization and absorbs neutrons with the opposite polarization, and (b) the initially considered multi-splitter-polarizer setup \cite{Zaliznyak_PhysicaB2005,ZaliznyakLee_MNSChapter}, where both the deflected and the transmitted neutron beams with the opposite polarizations are separated on the detector and measured simultaneously. The sample scattering volume accepted by the radial supermirror polarizer, (a), is determined by the supermirror critical angle and the distance to the focal spot at the sample position, (c1), while in the initially considered multi-polarizer setup, (b), the accepted sample size is determined by the width ($\sim 20$~mm) and the collimation ($\sim 20'$) of each polarizer channel, (c2). The radial polarizer effectively acts as a radial collimator, where the angular acceptance is determined by the supermirror critical angle and is energy-dependent, and so is the accepted sample size, (d). The solid line in (d) shows the effective Gaussian FWHM angular acceptance (right scale) and the sample size (left scale) corresponding to $4.5\theta_c^{Ni}$ (we assume m=3 on the outer and m=1.5 on the inner surface of each supermirror channel, in accordance with  \cite{Filges2006}); the upper broken line shows the square-window acceptance for $m=4.5$ and the lower broken line for $m=3$. The horizontal dotted lines illustrate typical values for $E_i = 20$~meV (vertical dotted line). The symbols show the FWHM of the beam profile at the sample position determined experimentally by scanning the translation of an incoherent scatterer (vertical plastic rod of 5mm diameter) in the beam.}
\label{fig:PolarizationAnalysis}
\end{figure}

The HYSPEC setup used for polarized beam measurements is shown in Figure~\ref{fig:InstrumentView}(a). A magnetically saturated Heusler (Cu$_2$MnAl) focusing crystal array mounted in a magnetic yoke, Fig.~\ref{fig:InstrumentView}(b), polarizes and vertically focuses the reflected neutron beam onto the sample. It is positioned in the drum shield in place of a pyrolytic graphite (PG) array, Fig.~\ref{fig:InstrumentView}(c), using a vertical translation stage. A sample, Fig.~\ref{fig:InstrumentView}(d), is mounted inside a variable-temperature environment, which is placed on a standard 2-axes goniometer inside a water-cooled coil system capable of providing an adiabatic magnetic guide field of up to $\sim 50$~Gauss. The direction of the guide field can be freely adjusted as required by experiment using the instrument control software. In particular, this allows a XYZ polarization analysis \cite{Scharpf1993,Ehlers_RSI2013,Ehlers2015}. With coils, the sample tilts are limited to $\approx \pm 5^\circ$.

For the un-polarized measurements, an $\approx 4.3^\circ$ radial collimator (RC, $0.67^\circ$ channel spacing) is placed at the entrance of the detector vessel, $\approx 53$~cm after the sample, Fig.~\ref{fig:InstrumentView}(e). For the polarized beam measurements, RC is replaced by the similarly positioned, magnetically saturated, yoked 60$^\circ$-wide-angle multi-channel radial supermirror-polarizer (RSP) array, Fig.~\ref{fig:InstrumentView}(f) \cite{Filges2006,PINS2006,Filges2012}. The calibration measurements performed using an incoherent scatterer (plastic, Vanadium, or TiZr alloy) show that while the transmission of the radial collimator is sufficiently flat as a function of scattering angle, the transmission of the RSP is rather inhomogeneous, with variations of up to $\pm \approx 40\%$. Although with the suitable calibration and correction procedure \cite{Savici_PNCMI2016} this does not present operational problems, an improved alignment procedure is presently discussed, which has the goal to overcome this behavior. The RSP transmission is also strongly energy dependent, decreasing from $\approx 40\%$ at 5 meV to about 10\% at 20 meV \cite{Savici_PNCMI2016}. Because polarization selection is achieved by reflection(s) from the supermirror, the RSP deflects the transmitted polarized beam by an angle, $\theta_{d}$, which is determined by the curvature of the supermirror channel, Fig.~\ref{fig:PolarizationAnalysis}(a). The latter has been optimized to maximize the transmission while still avoiding the direct view, and thus keeping the high polarization efficiency \cite{Filges2006,Filges2012}.
The curvature of the supermirror channel was calculated with the formula $R=L^2/8d$, where $L$ is the length of the analyzer (171 mm) and $d$ is the channel opening at the exit side (0.566 mm) plus the substrate thickness of 0.2 mm. Finally, the radius was adjusted to 460 cm.
Accounting for the scattered beam offset on the detector and the scattering angle dependence of the RSP transmission is carried out in MANTID \cite{MANTID} data preprocessing, by adjusting the detector efficiency for the corresponding events. The (optional) accounting for the energy-dependent angle-averaged RSP transmission,  $T \approx T_0 e^{- E_F/E_0}$ ($T_0 \approx 0.59 $, $E_0\approx 10.8$), is performed at a later stage, on the histogrammed data. A detailed description of the polarized data analysis procedures is given in the accompanying paper \cite{Savici_PNCMI2016}.

The size of the sample scattering volume visible to detectors is determined by the size of the focal spot of either RC, or RSP. The latter is defined by the effective collimation, $\delta \theta$, and the distance ($\approx 53$~cm) between the sample and the RC, or RSP entrance, respectively, Fig.~\ref{fig:PolarizationAnalysis}(c1). While in the case of a rather coarse RC the size of the focal spot is well matched to the nominal 4~cm wide (maximum) illuminated sample size as determined by the beam (guide) width, in the case of the RSP the focal spot size is much smaller and is also energy-dependent, Fig.~\ref{fig:PolarizationAnalysis}(d). This is because the effective collimation of the narrow channels of the RSP is controlled by the critical angle of the supermirror coating, which is small and depends on the neutron wavelength. The resulting reduced accepted sample scattering volume imposes considerable penalty in intensity for larger samples, and also requires careful alignment of the sample on the rotation axis, and co-alignment of the sample axis with that of the RSP. A misalignment of the sample with respect to the rotation axis could move it in/out of the RSP focal view as a function of the rotation angle resulting in a spurious intensity variation, a potential source of a systematic error. This problem is easily solved when the analyzer has an adjustable rotation as foreseen in the design.

Currently, the polarizer alignment mechanism, which is an important feature envisioned in the design, is not yet installed. With the default, fixed RSP alignment used at present there is an offset of $\approx 5$~mm ($0.5^\circ$) between the RSP focal spot and the sample rotation axis on HYSPEC.
%This design feature was introduced %by the manufacturer \cite{Filges2012} with the view of increasing the RSP polarization efficiency from $\approx 94\%$ to $\approx 95\%$ (note that this corresponds to $\approx 20\%$ lower probability for transmission of the undesired neutron polarization) at the cost of a $\approx 25\%$ intensity loss for $E_i \approx 6$~meV neutrons.
This improves the RSP polarization efficiency reducing the probability for transmission of the undesired neutron polarization by $\approx 20\%$ (the polarization efficiency increases from $\approx 94\%$ to $\approx 95\%$) at the cost of a similar, $\approx 25\%$ intensity loss for $E_i \approx 6$~meV neutrons \cite{Filges2012}. During the design phase of the RSP is was decided to concentrate on a high polarization mode at low $E_i$. Now we know, this offset is disadvantageous on HYSPEC, as it leads to a progressively larger loss of flux from the sample at higher $E_i$, and also introduces an ill-controlled systematic error associated with the rotation of a single crystal sample. The commissioning of the polarizer alignment mechanism, which would allow centering the RSP focal spot at the sample is planned for the near future.

The sample size limitations imposed by the RSP potentially could be alleviated in the beam-splitter \cite{ZaliznyakLee_MNSChapter} multi-analyzer setup, Fig.~\ref{fig:PolarizationAnalysis}(b),(c2), initially considered for HYSPEC \cite{Zaliznyak_PhysicaB2005}. However, the potential gain in sample size visible to detectors, which is determined by the width (2~cm) of a single channel, would be outweighted by the loss from the $20'$ collimation required in such a setup. There would also be significant disadvantages associated with the poorly controlled polarization mixing on detector, leading to a detector- and energy-dependent flipping ratio (FR), and a complicated measurement and calibration procedure. The follow-up detailed comparative studies and simulations of the different polarization analysis options \cite{Filges2006,PINS2006} have favored the radial supermirror polarizer implemented on HYSPEC \cite{Winn2015}.

\section{Polarization analysis}

\subsection{Spin-flip and non-spin-flip intensity measurement}
The HYSPEC polarizing supermirror array can be magnetized either up, or down using the dedicated electromagnet charging station. The default mode of operation is to have the RSP magnetized so that it transmits neutron polarization opposite to the polarization of the incident beam, which is reflected by the Heusler crystal array (magnetized horizontally) and adiabatically rotated to the vertical direction by a suitably coupled permanent magnetic guide field. The measurement with the Mezei spin-flipper in front of sample turned off then yields a spin-flip (SF) scattering intensity and the measurement with the flipper on yields a non-spin-flip (NSF) signal. With this choice, the flipper efficiency (which is very close to 1, \cite{Winn2015}) shows up in the NSF intensity and does not affect the SF intensity.

The finite efficiencies of the polarizing elements lead to a finite probability, $p$, for the non-spin-flip sample scattering intensity, $I_{NSF}$, to be detected with the flipper off and contribute to the measured spin-flip intensity, $I^{exp}_{SF} = pI_{NSF} + (1-p)I_{SF}$, and vice versa, for the $I_{SF}$ to contribute to the measured $I^{exp}_{NSF} = (1-p)I_{NSF} + pI_{SF}$. This probability is quantified by the experimentally measured flipping ratio, $FR$, which is determined as the ratio of the scattered intensities measured in the non spin flip, $I^{exp}_{NSF}$, and the spin flip, $I^{exp}_{SF}$, channels, or with the flipper ON and the flipper OFF, respectively, for a non-magnetic nuclear Bragg scattering, $I_{nuc}$. If we denote the instrument flipping ratio, $F = 1/p$, as an inverse probability, $p$, for a neutron scattered by sample in the NSF channel to be measured in the SF channel, and vice versa, then $I^{exp}_{SF} = pI_{NSF} = \frac{1}{F}I_{nuc}$, and $I^{exp}_{NSF} = (1-p)I_{NSF} = \frac{F-1}{F}I_{nuc}$,
\begin{equation}
FR=\frac{I^{exp}_{NSF}}{I^{exp}_{SF}} = \frac{1}{p} -1 = {F-1}.
\end{equation}
The corrected sample scattering intensities, $I_{SF}$ and $I_{NSF}$, are then obtained from the measured ones in a conventional way, by applying this measured flipping ratio correction,
\begin{eqnarray}
I_{SF} & = &\frac{1-p}{1-2p}{I^{exp}_{SF}} - \frac{p}{1-2p}{I^{exp}_{NSF}} \nonumber \\
& = &\frac{FR}{FR-1}{I^{exp}_{SF}} - \frac{1}{FR-1}{I^{exp}_{NSF}}, \nonumber \\
I_{NSF} & =  &\frac{1-p}{1-2p}{I^{exp}_{NSF}} - \frac{p}{1-2p}{I^{exp}_{SF}} \nonumber \\
& = &\frac{FR}{FR-1}{I^{exp}_{NSF}} - \frac{1}{FR-1}{I^{exp}_{SF}}.
\label{FRcorrection}
\end{eqnarray}

\subsection{The configuration of the guide field and $XYZ$ polarization analysis}
The neutron polarization at the sample position, ${\bf P}$, can be adiabatically guided to point in any predetermined direction with respect to the instrument coordinate frame using the guide field provided by the coil system, Fig.~\ref{fig:InstrumentView}(a), \cite{Winn2015}. In a conventional triple-axis setup with narrow, collimated beams, this uniquely determines the orientation of the polarization, ${\bf P}$, with respect to the wave vector transfer, ${\bf Q} = {\bf k}_i - {\bf k}_f$ (${\bf k}_i$ and ${\bf k}_f$ are the incident and the scattered neutron wave vectors), in the sample reciprocal space. In a wide-angle TOF setup, on the other hand, the detector bank registers neutrons with scattered wave vectors, ${\bf k}_f$, filling a large volume of phase space and having a wide range of directions and magnitudes. As a result, the direction of ${\bf Q}$ varies significantly as a function of the energy transfer and the detector pixel position, which defines the direction of the scattered neutron wave vector, ${\bf k}_f$. This direction is conveniently parameterized by two angles, the scattering angle in the horizontal plane, $S2$, between ${\bf k}_i$ and the horizontal projection of ${\bf k}_f$, and the vertical angle, $\eta_v$, of ${\bf k}_f$ with respect to the horizontal plane. S2 is determined by the position of a (vertical) detector tube in the detector bank and of the detector bank itself, while  $\eta_v$ is determined by the pixel position on the detector tube.

\begin{figure}[!t!h]
%\begin{center}
\includegraphics[width=0.49\textwidth]{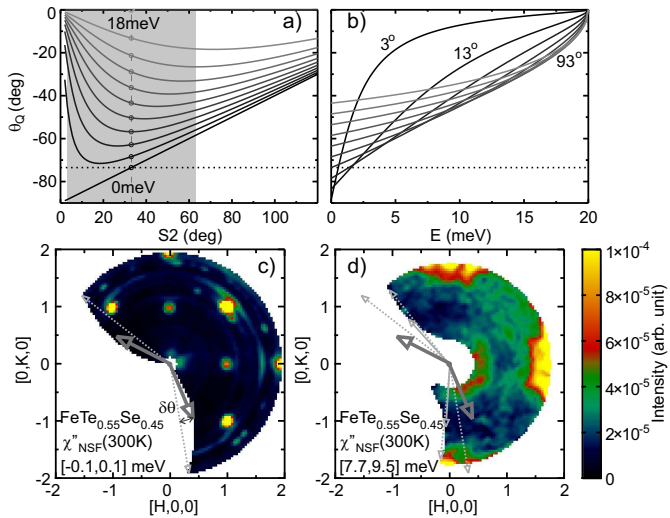}
%\end{center}
\caption{Dependence of the angle between the incident beam direction, {\mbox{\boldmath$k$}}$_i$, and the wave vector transfer, {\mbox{\boldmath$Q$}}, on the scattering angle in the horizontal plane, $S2$, for different energy transfers, (a), and on the energy transfer for different scattering angles, (b), for $E_i = 20$~meV. Curves in (a) correspond to energy transfers from 0~meV to 18~meV, with the step of 2 meV, from bottom to top. The shaded area shows the $60^{\circ}$ range of scattering angles accepted by the analyzer-detector system; the vertical dotted line and symbols correspond to the detector center, positioned at $S2 = 33^{\circ}$. Curves in (b) correspond to scattering angles from $3^\circ$ to 93$^\circ$, with the step of 10$^\circ$ (dark to light). In measurements with the horizontal guide field (HF) presented here, the neutron polarization, {\mbox{\boldmath$P$}}, which is determined by the guide field at the sample, is set parallel to the wave vector transfer at zero energy transfer for the scattering angle at the center of the detector, dotted horizontal line in (a), (b). This is illustrated by the thick solid grey arrows in (c), where the slice of the measured elastic intensity is shown (two arrows correspond to the limits of the sample rotation range). The dotted grey arrows show the respective elastic wave vector transfers at the detector edges. The same arrows are shown on the inelastic data slice in (d) together with the thin solid and dotted arrows, which illustrate the (rotated) wave vector transfer, {\mbox{\boldmath$Q$}}, at the same detector scattering angles for the energy transfer $E = 8.6(9)$~meV.}
\label{fig:PolarizationAngle}
\end{figure}

From wave vector conservation, it is easy to obtain the expression for the angle, $\theta_{\bf Q}$, between ${\bf Q}$ and ${\bf k}_i$ (the incident beam direction),
\begin{equation}
\sin{\theta_{\bf Q}} = -\frac{k_f}{Q} \sin{(2\theta_s)},
\label{theta_Q}
\end{equation}
where $2\theta_s$ is the sample scattering angle, i.e. the angle between ${\bf k}_i$ and  ${\bf k}_f$ in the scattering plane. In general, the scattering plane, which is defined by these two wave vectors is not horizontal (for out-of-plane detector pixels), but is at an angle $\alpha$ to the horizontal plane, where
\begin{equation}
\sin{\alpha} = \frac{\sin{\eta_v}}{\sin{(2\theta_s)}}, \;\;\;\; \cos{\alpha} = \frac{\sin (S2)}{\sin{(2\theta_s)}}\cos{\eta_v} = \frac{\tan{(S2)}}{\tan{(2\theta_s)}}.
\end{equation}
The scattering angle is given by,
\begin{equation}
\cos{(2\theta_s)} = \cos{(S2)} \cos{\eta_v},
\end{equation}
so $2\theta_s > S2$ for $S2 < 90^\circ$. For small scattering angles, the scattering plane can be strongly tilted even if $\eta_v$ is small, e.g. for small $\eta_v \approx S2$, $\alpha \approx 45^\circ$. The angle with respect to the horizontal plane for the wave vector transfer, $\eta_{\bf Q}$, is given by a relation similar to Eq.~(\ref{theta_Q}),
\begin{equation}
\sin{\eta_{\bf Q}} = -\frac{k_f}{Q} \sin{\eta_v} .
\label{eta_Q}
\end{equation}
Even though detector vertical acceptance on HYSPEC is relatively small,  $\eta_{v} < 7.5^\circ$, for small wave vector transfers, $Q < k_f$, $\cos{(2\theta_s)} > \frac{k_i}{2k_f}$, $\eta_{\bf Q}$ can be large. For neutron energy-loss scattering, $k_i \geq k_f$, $\left( \frac{Q}{k_f} \right)^2 = \left( \frac{k_i}{k_f}-1 \right)^2 + 4\frac{k_i}{k_f} \sin^2{\theta_s} \geq 4 \sin^2{\theta_s}$, and an upper bound for $\eta_{\bf Q}$ is given by,
\begin{equation}
\sin{\eta_{\bf Q}} \leq \frac{\sin{\eta_v}}{2\sin{\theta_s}}.  %= \frac{\sin{\eta_v}}{\sqrt{2 - 2\cos{(S2)} \cos{\eta_v}}} \leq \sin{\alpha}
\label{eta_Q_bound}
\end{equation}
Here the equality corresponds to elastic scattering, where $\eta_{\bf Q}$ is the largest. For the top and the bottom detector pixels on HYSPEC, $\eta_v \approx 7.5^\circ$, this gives (in the elastic channel) $\eta_{\bf Q}$ of $\approx 68^\circ$, $\approx 23^\circ$, $\approx 13^\circ$ and $\approx 7^\circ$ for the horizontal scattering angle, $S2$, of $3^\circ$, $18^\circ$, $33^\circ$ and $63^\circ$, respectively. The increase of $\eta_{\bf Q}$ with $\eta_v$ is most pronounced at low angles, $S2 \leq 2 \eta_v$. Only for $\theta_s \geq 30^\circ$ ($S2 \geq 60^\circ$), where $\sin{\eta_{\bf Q}} \leq \sin{\eta_v}$ holds, $\eta_{\bf Q}$ can be safely neglected if $\eta_v$ is small.

What the above considerations show, is that the direction of ${\bf Q}$ varies strongly with respect to the plane defined by the two vectors, ${\bf k}_i$ and ${\bf P}$, whether it is horizontal (${\bf P} \perp Z$), or vertical (${\bf P} \| Z$). This variation is well beyond the paraxial approximation na\"{i}vely implied by small $\eta_v$, and is especially dramatic at small $S2$ angles. It presents a challenge for the polarization analysis on HYSPEC, where magnetic scattering is routinely measured down to $S2 \leq 3^\circ$. In general, the $XYZ$ polarization analysis for a strongly out-of-plane scattering is quite complicated and requires one to account for the variation of the angle ${\eta_{\bf Q}}$ as a function of detector pixel and time (${\bf Q}$ and $E$). Methods for performing the $XYZ$ NPA in such cases were considered in Refs. \cite{Scharpf1993,Ehlers_RSI2013,Ehlers2015}. The advanced procedures for full NPA on HYSPEC are currently in development.

In the case of a coherent scattering in the horizontal plane, such as Bragg diffraction, the out-of-plane angular deviation of the scattered neutron wave vector is determined by the vertical resolution, and is $\eta_v \leq \pm 2.5^\circ$ on HYSPEC. The assumption that ${\bf Q}$ is in the horizontal scattering plane, then typically constitutes a reasonable approximation; the deviations must be treated in the resolution correction formalism for NPA, which needs to be developed. The dependence of the angle between ${\bf Q}$ and the incident beam direction (Eq.~\ref{theta_Q}) for scattering in the horizontal plane is illustrated in Fig.~\ref{fig:PolarizationAngle}. It shows $\theta_{\bf Q}$, as a function of the scattering angle, $S2$, for a setup with $E_i = 20$~meV, which was used in the polarized inelastic measurements on FeTe$_{0.55}$Se$_{0.45}$ presented below (Fig.~\ref{fig:FeTeSe_inelastic}).

For the full $XYZ$ NPA, at least 3 orientations of neutron polarization on sample have to be measured \cite{Scharpf1993,Ehlers_RSI2013,Ehlers2015}. In a polarized neutron experiment on a TAS with narrow, collimated beams, this is usually accomplished by performing measurements using the vertical guide field (VF, ${\bf P} \| Z$) and two mutually perpendicular horizontal field (HF) directions: along the wave vector transfer (${\bf P} \| {\bf Q}$) and pependicular to it (${\bf P} \perp {\bf Q}$). In the initial polarized measurements on HYSPEC reported here, we followed a similar procedure. The guide field defining the neutron polarization ${\bf P}$ at the sample was either vertical (VF), or horizontal (HF) and such that condition ${\bf P} \| {\bf Q}$ was satisfied for elastic scattering at the center of the detector bank (we still call this configuration ${\bf P} \| {\bf Q}$). According to Eq.~\ref{theta_Q}, in the case of elastic scattering ($k_i=k_f$, $Q = 2k_f \sin{\theta_s}$),
\begin{equation}
\theta_{\bf Q} = {\theta_s} - 90^\circ,
\label{HparQ}
\end{equation}
which also follows from the scattering geometry. In our measurements, the detector bank was positioned to span horizontal scattering angles $S2$ in the $[-63^\circ, -3^\circ]$ range, and HF (${\bf P}$) was directed at $\theta_{\bf Q} = -106.5^\circ$ (Fig.~\ref{fig:PolarizationAngle}, (a,b) illustrates the opposite scattering side, where $S2$ is in the $[3^\circ, 63^\circ]$ range and $\theta_{\bf Q} = -73.5^\circ$). Although the deviation of ${\bf P}$ from the ${\bf Q}$ direction as a function of energy transfer ($E_f$) is large, it is only $\pm 15^\circ$ across the whole detector bank for the elastic scattering, and even less than that for the angular range where Bragg reflections are observed (Fig. \ref{fig:YbMnBi2_diffraction}). Hence, neglecting the deviation of $\theta_{\bf Q}$ in the elastic channel is reasonably accurate (to within $\sim 10\%$). The wave vector range spanned by the detector bank and its rotation between the elastic and the inelastic ($E \approx 8.6$~meV) slices of the typical HYSPEC data measured in the polarized mode with $E_i = 20$~meV are illustrated in Fig.~\ref{fig:PolarizationAngle}, (c) and (d), respectively.

\section{Polarized neutron diffraction in YbMnBi$_2$}

\begin{figure}[!t!h]
\vspace{-0.1in}
\includegraphics[width=0.49\textwidth]{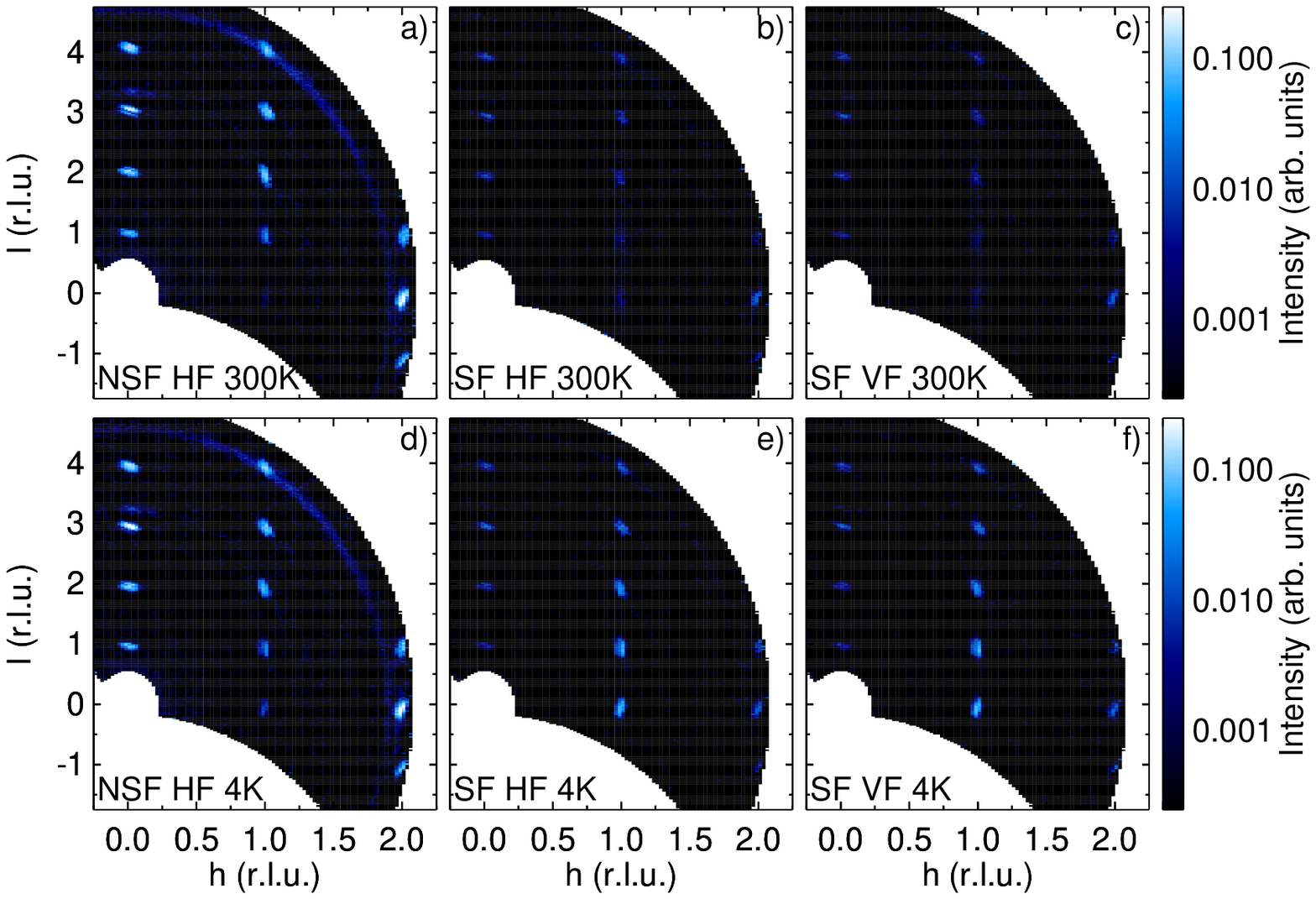}
\vspace{-0.1in}
\includegraphics[width=0.49\textwidth]{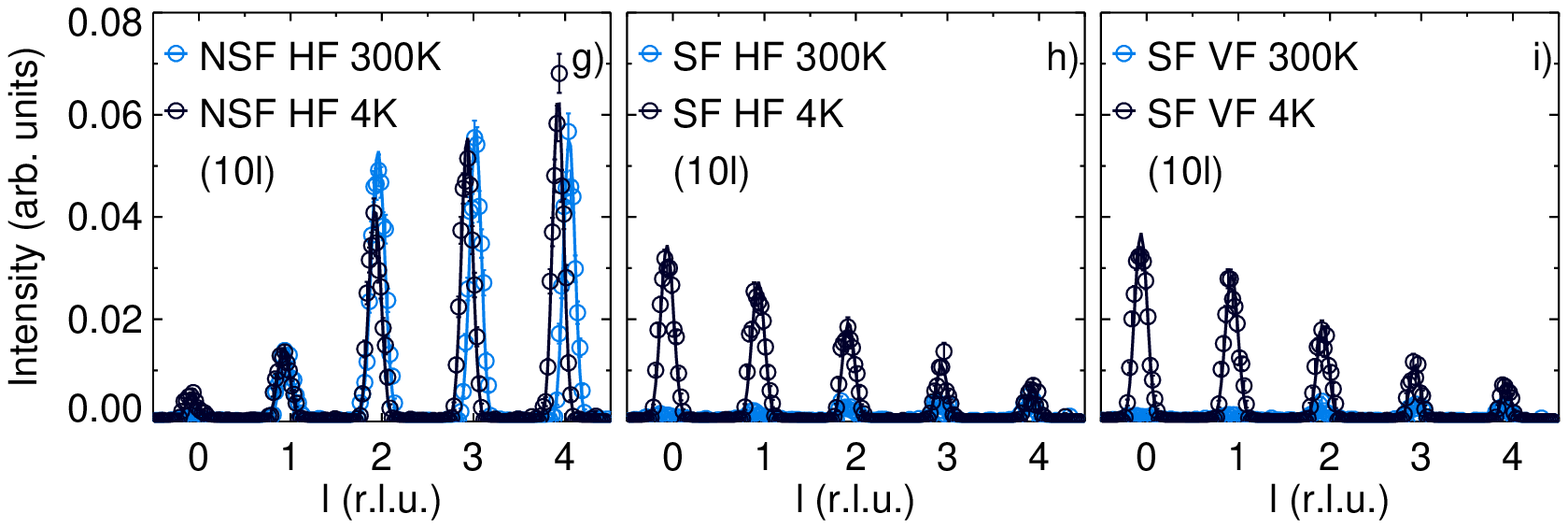}
\caption{
Polarized neutron diffraction from the YbMnBi$_2$ ($\approx 0.76$~g) sample in the $(h,0,l)$ plane measured on HYSPEC with $E_i = 15$~meV and $\nu_{Fermi} = 120$~Hz. The data was collected by rotating the sample with a $0.5^{\circ}$ step within the range of $130^{\circ}$, with about 25~s counting time at each angle setting. The measurement was repeated in the non-spin-flip (NSF, flipper on) and the spin-flip (SF, flipper off) channels with the horizontal guide field (HF) at the sample, at 300 K (a,b) and at 4 K (d,e), and in the SF channel with the vertical guide field (VF), also at 300 K (c) and 4 K (f). Apart from a small shift in position resulting from the temperature-dependent lattice parameters, there is essentially no change between 300 K and 4 K for $(0,0,l)$ peaks, and for $(1,0,l)$ peaks in the NSF channel, (g), where the nuclear lattice scattering dominates. There is, however, significant intensity arising at 4 K at the $(1,0,l)$ lattice positions, which shows up in the spin-flip channel, (h,i), manifesting the appearance of the antiferromagnetic order. } %The total collection time for all data shown in the figure was about 16 hours.
\label{fig:YbMnBi2_diffraction}
\end{figure}

Figure \ref{fig:YbMnBi2_diffraction} presents the results of the polarized neutron diffraction measurements on a candidate Weyl metal YbMnBi$_2$ \cite{Wang_etal_Petrovic_arXiv2016}. Measurements were performed on a 0.8~g high quality single crystal, with $E_i = 15$~meV neutrons and a Fermi chopper at 120~Hz. Each guide field/flipper configuration required about 100 min. counting time for $130^\circ$ crystal rotation. The nuclear Bragg intensities measured in a magnetically disordered phase at $300$~K indicate a flipping ratio $FR = 14(1)$, consistent for all guide field directions and for different Bragg peaks (sample rotation angles).

YbMnBi$_2$ crystallizes in a tetragonal (P4/nmm) symmetry, with alternating layers of square nets of Bi atoms hosting Dirac-like charge carriers, and square-lattice magnetic layers of Mn$^{2+}$, with 2 Mn per unit cell, structurally similar to those in cuprates and Fe-based superconductors, and with Yb spacer layers. YbMnBi$_2$ orders antiferromagnetically below $\sim 300$~K. Antiferromagnetic and crystallographic unit cells are identical, and therefore magnetic and nuclear Bragg positions coincide. However, magnetic and nuclear scattering can be separated using neutron polarization analysis at HYSPEC, as shown in Fig.~\ref{fig:YbMnBi2_diffraction}. Measurements confirm magnetic structure determined with unpolarized neutrons and the refined Mn magnetic moment of $4.3(1)\mu_B$, oriented along the $c$-axis.

\section{Inelastic magnetic and non-magnetic scattering in FeTe$_{1-x}$Se$_{x}$}

\begin{figure}[!h]
%\begin{center}
\includegraphics[width=0.49\textwidth]{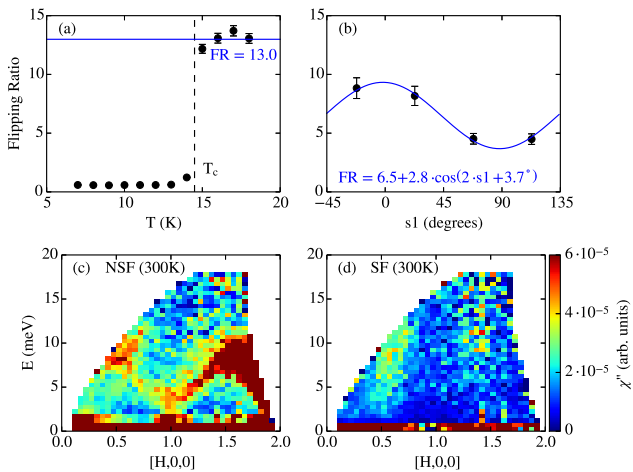}
%\end{center}
\caption{\label{fig:FR_SF_NSF_data} Data from the polarized inelastic measurement on HYSPEC ($ E_i = 20$~meV, $\nu_{F} = 120$~Hz) of a superconducting FeTe$_{0.55}$Se$_{0.45}$ single crystal ($m=23.4$~g, mosaic $\approx 2.2^{\circ}$ FWHM) in the $(h,k,0)$ zone. (a) Bragg flipping ratio measured on warming after the sample was cooled through the superconducting transition temperature, ${T_c}$, in a horizontal guide field of $\approx 20$~Gauss. (b) Bragg flipping ratio at $T = 5$~K in the same horizontal guide field ($\approx 20$~G), as a function of sample rotation angle after the sample is cooled in zero magnetic field. The imaginary part of the dynamical susceptibility obtained from the non-spin-flip (NSF), (c), and the spin-flip (SF), (d) intensity measured at $T = 300$~K by correcting for energy-dependent polarizer transmission and the detailed balance factor.}
\label{fig:FeTeSe_inelastic}
\end{figure}

Magnetic excitations in the parent materials of the chalcogenide family, Fe$_{1+y}$Te, and in the superconducting versions of it obtained by partial substitution of Te by Se, or S, reveal rather unusual behavior, which does not comply with the standard spin-wave description. In particular, dynamical magnetism increases in intensity with increasing temperature \cite{Zaliznyak_PRL2011}, a rare and unexpected phenomenon. This unusual temperature dependence of magnetic intensity suggests a non-trivial interaction between local spins, which give rise to the bulk of the spectral intensity, and the itinerant band electrons, which also contribute to it. So far, scatering patterns were identified as magnetic based on their ${\bf Q}$-dependence, but magnetic and lattice contributions can only be distinguished using polarized neutrons. The maxima in the energy dependence of magnetic intensity occur near the top of the acoustic phonon dispersion ($\sim 8$~meV), at the exact positions where diffuse columns of magnetic scattering meet the phonons. Very similar observations were made in superconducting FeTe$_{1-x}$(S,Se)$_{x}$ samples \cite{Zaliznyak_PNAS2015,Xu_etal_PRB2016}, where the temperature-induced dynamical magnetism was also observed.

The magnetism and the lattice in iron chalcogenides are closely coupled. A bond-order wave (BOW) transition leading to ferro-orbital order in the ground state associated with formation of zigzag Fe-Fe chains stabilizes the bicollinear antiferromagnetism \cite{FobesZaliznyak_PRL2014}. This latter transition entails lowering of the lattice symmetry so that structure factor for the (1,0,0) lattice Bragg reflection, which is forbidden at high T, becomes non-zero. Surprisingly, an acoustic-phonon-like dispersion, centered at (1,0,0), is observed in the inelastic spectra in both the monoclinic P21/m and the tetragonal P4/nmm structural phases \cite{FobesZaliznyak_PRB2016}. This mode is unexpected because although in the monoclinic phase the phonon mode is accompanied by a Bragg peak at ${\bf Q} = (1,0,0)$, in the tetragonal phase elastic Bragg scattering at this ${\bf Q}$ is forbidden by unit cell symmetry and no accompanying Bragg peak was observed despite the continued observation of an acoustic phonon mode.

Using polarized neutrons on HYSPEC enables us to ascertain which scattering is magnetic or structural in origin, and allows to understand the interaction between electronic magnetism and the lattice degrees of freedom in this family of materials. Figure \ref{fig:FeTeSe_inelastic} presents the results of the polarized neutron inelastic scattering measurements on a single crystal sample [shown in Fig.~\ref{fig:InstrumentView}(d)] of an optimally doped 11 iron chalcogenide superconductor FeTe$_{0.55}$Se$_{0.45}$ \cite{Xu_etal_PRB2016}. After the sample has been cooled through the superconducting transition at $T_c \approx 14$~K in a 20~Gauss guide field, the flipping ratio deteriorated, because the neutron beam was completely depolarized by small random magnetic fields frozen inside the sample, Fig.~\ref{fig:FeTeSe_inelastic}(a). This problem was largely resolved by cooling the sample in a zero field environment. Both placing the cryostat in a $\mu$-metal shield, and on the sample table and using the compensating field created by the coil system, lead to similar results shown in Fig.~\ref{fig:FeTeSe_inelastic}(b). After the zero-field cooling, the flipping ratio varied between $\approx 4$ and $\approx 8$, as a function of the sample rotation angle, $S1$. This is understandable, because the large and irregularly shaped superconducting sample still somewhat depolarizes the neutron beam, and this depolarization depends on the sample orientation with respect to the guide field. Symmetry suggests that this dependence should be $180^\circ$-periodic as a function of $S1$, $FR = FR_0 + f \cos{(2S1 + \phi_0)}$, Fig.~\ref{fig:FeTeSe_inelastic}(b).

The flipping ratios at several different $S1$ were obtained from the SF and NSF intensities of the nuclear Bragg peaks, which were measured in the same experiment as the inelastic data [see Fig.~\ref{fig:PolarizationAngle}(c)]. Even though $FR$ is not very high and angle-dependent, the data can be reliably corrected using Eq.~(\ref{FRcorrection}) and the fitted angular dependence of $FR$. The flipping ratio for the 300~K data presented in Fig.~\ref{fig:FeTeSe_inelastic}(c),(d), on the other hand, was $13(1)$, independent of $S1$. The measured NSF intensity [Fig.~\ref{fig:FeTeSe_inelastic}(c)] clearly reveals the acoustic phonon mode coming out of the forbidden $(1,0,0)$ Bragg peak, joining at $\approx 8$~meV with some NSF intensity at small angles, which is probably of magnetic origin. The SF intensity [Fig.~\ref{fig:FeTeSe_inelastic}(d)] reveals columns of magnetic scattering above a gap of $\approx 5$~meV, at ${\bf Q} = (0.5, 0, 0)$ and $(1.5, 0, 0)$.

\section{Summary and conclusions}
This paper describes the instrument setup, and procedures for the polarized beam measurements on the HYbrid SPECtrometer at the SNS. We also present the first results of the diffraction and the inelastic measurements with the polarization analysis. Despite the significant penalty in intensity imposed by the lower (compared to PG) reflectivity of Heusler crystals, and low transmission of the supermirror-array polarization analyzer, broad surveys of the scattered SF and NSF intensities are possible for incident neutron energies as high as $20$~meV. For non-ferromagnetic and non-superconducting samples the flipping ratio is high, in the range of 13 to 15. However, broad polarized beam surveys, such as performed on HYSPEC, can tolerate situations where it is reduced. In an experiment with sample rotation, the flipping ratio can be measured on several nuclear Bragg peaks throughout the sample rotation range, which allows reliable FR correction using Eq.~(\ref{FRcorrection}). While distinguishing SF and NSF scattering is straightforward, the full $XYZ$ polarization analysis requires a complex procedure accounting for the change in ${\bf Q}$ orientation as a function of time of flight and detector position, which is still in development. Another, related challenge is in devising a resolution correction procedure for the $XYZ$ NPA, which accounts for the appreciable vertical spread of the focussed beam, whose impact is especially severe at small angles.

\section{Acknowledgements}
Authors are grateful to M. Hagen for his leadership role in HYSPEC construction, to M. Kenzelmann for valuable discussions, and to M. Graves-Brooks for help with the measurements. This work was supported by the Office of Basic Energy Sciences (BES), Division of Materials Sciences and Engineering, U.S. Department of Energy (DOE), under Contract No. DE-SC00112704.
Research conducted at ORNL's Spallation Neutron Source and High Flux Isotope Reactor was sponsored by the Scientific User Facilities Division, Office of Basic Energy Sciences, US Department of Energy.
This work was supported by the Paul Scherrer Institut by providing the supermirror analyzer as a temporary loan to Oak Ridge National Laboratory.
This manuscript has been co-authored by UT-Battelle, LLC under Contract No. DE-AC05-00OR22725 with the U.S. Department of Energy. The United States Government retains and the publisher, by accepting the article for publication, acknowledges that the United States Government retains a non-exclusive, paid-up, irrevocable, worldwide license to publish or reproduce the published form of this manuscript, or allow others to do so, for United States Government purposes. The Department of Energy will provide public access to these results of federally sponsored research in accordance with the DOE Public Access Plan (http://energy.gov/downloads/doe-public-access-plan).

%\bibliography{HYSPECbib}

\providecommand{\newblock}{}

\end{document}